# Building the distributed WPS-services execution environment

Bychkov V. I.[1], Ruzhnikov M. G. [2], Fedorov K. R. [3], Shumilov S. A. [4]

[1]*Institute for System Dynamics and Control Theory*

*of Siberian Branch of Russian Academy of Sciences,*

*Irkutsk, Russia, bychkov@icc.ru, +73952 427100*

[2]*ISDCT SB RAS, Irkutsk, Russia, rugnikov@icc.ru, +73952 453108*

[3] *ISDCT SB RAS, Irkutsk, Russia, fedorov@icc.ru, +79501320841*

[4] *ISDCT SB RAS, Irkutsk, Russia, alexshumilov@yahoo.com, +79501017750*

**Abstract**

The article describes the environment of WPS-based (Web Processing Service) distributed services, that uses scenarios in JavaScript programming language in order to integrate services with each other. The environment standardizes data processing procedures, stores all services-related information and offers the set of basic WPS-services.

**Keywords**

OGC, WPS, JavaScript, V8, SOA

**Introduction**

Usually the solution of any complex scientific task requires the application of a certain number of programming systems that are developed by specialists from the different fields of science. For example, calculation of air pollution for certain geographic region uses methods of controlling pollution sources (roads, heaters, power plants) and modeling of pollution spread. These methods are commonly complex and usually are developed by different teams. Collaboration between teams is rare because of the complexity of the methods usage and difficulties of different software packages integration. As a result, one or more subtasks are solved at the proper level depending on developer's team resources, and the rest of tasks are solved by simplifying and limiting initial requirements.

Steady increase of average Internet speed, development and standardization of web-browser's programming interfaces allow a lot of programming systems to be implemented as web-services, accessed through the HTTP protocol. Implementation as a web-service results in following benefits – scalability, easier distribution, centralized updates, usage of shared information resources etc. [1]. A lot of formerly desktop-only applications are migrating to web-sphere, such as Abobe Photoshop™, Microsoft Office™, ArcGIS™. Web-services allow cooperative work on documents, maps, media etc.

Interoperability between information systems and software packages through Internet, based on OGC (Open Geospatial Consortium) [2] standards, is actively developed as well. One of the most interesting standards for this work is OGC WPS (Web Processing Service) [3] standard, which unifies the way clients interact with web-services that deal with geospatial data analysis and processing. For instance, it could be the service for raster and vector data processing, service for geo modeling and statistics. This standard is simple, it defines metadata distribution and supports long-lasting service execution.

Development of modern information technologies, as well as constant development of standards, allows establishment of collaboration of specialists from different fields of study at a whole new level using service-oriented architecture. The article describes the way distributed data and WPS-services can be organized into the solid programming systems to solve complex tasks.

**Related work**

Further some of the currently developed WPS-based projects are listed:

- deegree WPS [4]. Project allows publishing of Java algorithms corresponding to certain standards using degree classes and modules, created by developers;

- ZOO Project WPS [5]. Algorithms, written in different languages (C++, Java, Python, Perl etc.) can be published and linked in complex structures using JavaScript language;

- GeoServer WPS [6]. This project does not allow publication of any service by developer, however, it combines variety of such OGC standards, as WMS, WFS and WCS, also it is oriented to the publication of geospatial data.

- pyWPS [7]. Project allows publication of services that are written in Python language.

- 52° North WPS [8]. Supports algorithms from different GIS: GRASS GIS, Sextante, ArcGIS.

Currently there are a few methods of solution of complex problems using combinations of WPS-services. Xiaoliang Meng, Yichun Xie, Fuling Bian define three ways of solving complex problems in their article Distributed Geospatial Analysis through Web Processing Service: A Case Study of Earthquake Disaster Assessment [9]. First method uses BPEL for building execution chain of services. BPEL [10] – XML-based standard, that defines way of description and execution of tasks in distributed environment. This method does not define the way how binary data is transferred between services, also it requires advanced knowledge of XML standard. Second method is realization of service chain as a standalone application that realizes the WPS standard, so this application will be compatible with other WPS-services. Third method is implementation of the service chain as a sequence of GET requests with processing of XML responses. Most interesting methods in the context of the current work are methods that use BPEL and WPS-compatible application. The method, proposed in this article, takes advantages from both methods that were listed above – variety of available logical constructions in JavaScript scenarios (comparable to BPEL) and ability to publish scenarios as regular WPS-services.

One of the most important parameters of the distributed WPS-services environment is its ability to maintain reusability of data. It means that any data that is a result of WPS-service can be used as an input to any other WPS-service. The example of such environment is described in the article Towards The Transactional Web Processing Service [11] by Bastian Schaeffer, where scenarios are written using BPEL standard and input parameters for services are defined with XML Schema standard. The author of the article proposes to treat any BPEL scenario as a common WPS-service, keeping the reusability of data at the same time. As an example, the author describes the service, that calculates air pollution and is implemented with 52° North WPS platform. This article, besides the definition of reusability of data, proposes powerful tool for input data validation – XML Schema standard.

Talking about distributed data processing using WPS-services, the article Grid Computing Enabled Web Processing Service[12] by Bastian Baranski should be mentioned. This article focuses on gridification problem – building the system that allows distributed processing of any WPS-service. Author proposes two ways of building such a system – using separate scripts for parallel WPS-service execution and actual integration of WPS-services in special software for distributed computing. Parallel WPS-service execution will definitely be needed in distributed environment of WPS-services, so the listed above article will, undoubtedly, be very useful in the future development.

Current limitations of WPS standard are listed below:

- Most of the WPS JavaScript clients implementations support WPS-services that are located on a single host, due to the JavaScript security limitations.

- Requirements for input parameters of WPS-services are too general, so they need more detailed specification. For example, if input parameter is the table of certain structure, where different types of data occur (coordinates, floats with certain number of decimal digits), then filling and validation of such type of data will be hard to implement.

- Sometimes WPS-services use files that are available through the HTTP or by using WMS, WMF, or WMC protocols. Usually files can be accessed in read-only mode. WPS-services do not ensure the reusability of data, because WPS-service results cannot be used as an input for another WPS-service. So, user needs to publish the results of WPS-service execution, so they can be used lately using listed methods that are listed above. Automatic publication is missing. So, processing the data, that is an output and input at the same time, is hard to implement.

- WPS-services that use files as an input data add some complexity. Files to be transmitted to the WPS-service are referenced by URL link, that can be used to fetch them through the HTTP protocol. File-access policy is not regulated by the standard, so it has to be open.

## WPS-service environment organization

As a result of analysis of current theoretical researches and software realization of different approaches, as well as investigation of pros and cons of WPS standard and distributed process execution in general, it is clear that there is a strong need in the development of distributed environment of WPS-services. The distributed environment of WPS-services has to be able to store user data and ensure that any WPS-service can perform operations over this data. In order to organize WPS-services execution the system has to be able to collect and store information about services, control their execution, provide input data and fetch result of WPS-services execution.

According to requirements that are determined to the execution scheme of WPS services, following a brief mathematical model was composed. In the simplest case the description of WPS services execution can be presented in a form of the following structure $I = (D, T, type, in, out, S)$, where

$D$ – the set of any possible input and output data for services,

$T$ – the set of data types that can be processed by WPS services execution scheme,

$S$ – the set of services that perform data manipulations,

$in$ – the set of any possible input data for specific service s, $in(s) \in D$,

$out$ – the set of any possible output data for specific service s, $out(s) \in D$,

$type(d, s)$ – the function that returns the data type of parameter $d$ according to possible data types of the service $s$.

The set of processed data types contains simple (textual and numeric data) and complex data types (raster and vector data with corresponding projection, other data types that can be defined with three characteristics according to the WPS standard – mime-code, encoding and schema).

It is essential for service execution to process input and output service parameters correctly in order to establish the data transmission between services, so the processed data can be reusable, i.e. the following condition has to be right: $\forall s1 \in S, s2 \in S: \forall a \in out(s1): \rightarrow a \in in(s)$ if $type(a, s1) = type(a, s2)$, so the closeness is introduced – any resulting data of the service can be the input data for another service if theirs data types match.

In order to realize formal model of the execution of WPS-services it is essential to create user data storage, inventory of WPS services, execution of WPS-services over the data, transmission and reception of data, WPS-services execution control.

**Storing the environment data**

One of the most important features of the distributed environment of WPS-services is provision of input data and collection of resulting data of service execution over the Internet, so special data storage systems are required in order to provide not only read, but also a write access mode for datasets. Most of the times files and data from relational database are the input data for WPS-services. WPS-services require support of the following protocols and data access standards: HTTP, WFS, WFS-T, WCS, WMS. Considering that sometimes files have big size, use of WFS and WFS-T becomes problematic because of textual representation of data. WMS and WCS standards does not allow altering of the data. Those are reasons why support for HTTP and Simple Features Specification For SQL has to be implemented in the first turn. Existing Internet file systems allow reading of the data, but data altering requires authentication, which is not the feature of WPS standard. This is why certain components for data processing are required in order to ensure data access according to WPS standard. Following component were developed as parts of the environment:

1) file data storage system where certain amount of disk space is given to every user of the system and where file access is standardized according to the WPS;
2) PostgreSQL (with PostGIS extension) for storing relation data, separate schema is created for every user as well;

File manager and interface for editing and displaying relational data is provided through the web-browser.

## WPS-services catalog

In order to organize WPS-services execution it is essential to know theirs distribution over the Internet, so tools for searching and registration of WPS-service has to be developed. Also, more specific configuration and description of input and output data has to be implemented. Detailed description allows to simplify the user-interface and validate input data before it will be submitted to the actual WPS-service. Search and registration of WPS-services is implemented in developed WPS service catalog. The catalog module is implemented as a module of CMS Calypso [13]. Registration of the WPS-service takes several steps. At first step user enters name of the service, its description, defines parameters of the remote WPS-platform. At the second step catalog queries the specified WPS-platform in order to fetch the list of available WPS-services. After user selects any of the available services, catalog queries the WPS-platform in order to get detailed information for the selected service. At the last step user defines input widgets for every input parameter of the service, fills the human-readable name and description of the parameter. This information is used for the rendering of the user-interface and at the validation of input data according to policies defined in corresponding widgets. Catalog allows to perform search of methods based on service name and description. Service type and input / output parameters search is being developed currently.

Before the execution of any WPS service, is it necessary to ensure that the user-friendly interface will simplify selection of data from data storage system or relational database as input or output parameters. That is the reason why widgets, that were selected at the service registration step, are used at the generation of the execution form. Widgets are used to simplify, clean or modify input data. For example, parameter "Source" has "file" widget assigned to it, so this widget will show data storage system interface, so user will be able to select file as an input parameter of the service.

Following widgets were developed: "edit" – for filling simple line values, "number" – for filling number values, "checkbox" – for entering boolean parameters, "rectangle" – for entering extent (rectangular area on the map) etc. Some widgets were developed specially for interaction with data storage system and database: "file" – for selecting file from data storage system, "select_table" – for selecting table from the database, "select_table_attr" – for selecting attribute of the specific table etc. When widget "file", "file_save" or "rectangle" are selected for any of the service parameters, interactive map is displayed inside of the service execution form.

**WPS-services management subsystem**

After the submission of data in the web-form, the environment of distributed WPS-services has to launch service, ensure data transmission and reception according to WPS standard and defined security policies. Execution form of the WPS service forms set of parameter names, then they are forwarded to the execution subsystem, the execution subsystem processes parameters in order to make them compatible with WPS standard, registers instance of executed service, performs Execute request to the service and fetches results of its work. Input parameter values are processed according to parameter type:

1. **File.** If transmitted parameter is a file, that is located in user folder in the data storage system, then it needs to be passed to the WPS-service. According to the WPS-standard, if the service has a file as an input parameter, then this parameter has to be passed as the URL that points to the web-accessible file, so WPS-service will be able to download this file by itself. At the same time, the provided URL can result in download of the file with confidential data, so it has to be a mechanism that allows download of the file only by WPS-service, at the same time it has to prevent download by unwanted subjects. Following solutions for this problem was found – every WPS-service instance has unique identification, so for every file parameter of current WPS-service the unique assigned to the instance id link is generated. Using the generated link, the service can download the file certain number of times. When service instance is finished, all generated links are terminated, so they can not be reused. If one of service results is a file, then execution system fetches the link from the service response, downloads the file that is pointed by the link and puts file at predefined path at the local data storage system.

2. **Extent** (rectangular area on the map). Client-side processes extent in WKT [14] format, execution module transforms it according to the WPS standard at the server-side.

3. **PostgreSQL database**. When database table has to be transmitted as a parameter for WPS service, GDAL [15] database connection string is formed (GDAL database connection string was used in order to provide security, because it does not expose any data to end-user), so the actual WPS-service will use it to connect to the database. The current version allows passing tables to local services only, but the next release will allow connection to the database for any of the WPS-services.

**WPS-services execution**

Available nowadays WPS-service execution methods offer execution of a single service or the sequence of services. For instance, Business Process Execution Language (an XML-based standard of describing process workflow) can be used for creating complex scenarios for solving variety of tasks [16]. However, solution of a non-trivial problem can require complex integration of distributed WPS services, which can involve parameter processing, WPS service iterations, conditions depending of middle results, etc. That is why the mechanism, which will allow to implement complex logic of the WPS-service scenario, is needed. One of the most flexible ways of doing this is to use existing programming language, which can realize any logic contained in the scenario. The JavaScript language was chosen for this purpose because of its prevalence. JavaScript will allow to execute WPS-services in the parallel mode because of it's ability to keep execution context and asynchronous execution, also it can support long-lasting execution of some of the WPS-services.

In order to implement complex logic of WPS-service application the special interpreter of WPS scenarios was developed. Before the WPS-service or scenario execution the code of this service or scenario is passed to the interpreter. Requests to WPS-service are performed by calling special JavaScript wrappers of WPS-services, registered in the system. JavaScript wrappers are automatically created for every WPS-service that is registered in the catalog. There is a special component for online scenario-editing, where user defines input and output parameters of the scenario, then writes actual scenario code. Besides standard scenario-editing form, component lists all available services and their wrappers, so they can be used in the scenario. When scenario is saved, it receives a identificator, unique wrapper name and is published as WPS-service.

**WPS-service scenarios interpreter**

Usage of the server-side interpreter provides two main advantages from the client-side execution in web-browser windows – it can freely connect to any hosts in order to execute WPS-service or query service data (if requests would be performed in web-browser, they would be limited by the security limitations) and does not depend on possible faults that can occur with the web-browser on the client-side – browser crash or the lack of resources. Service execution subsystem supports long-lasting services (for example, services that last few days) according to the WPS standard. When long-lasting service is executed, requests for its state and percent of completion are periodically sent, so the user's web-interface can display relevant information about the execution state. If the interpreter is integrated right in the WPS-application, that serves local WPS-services, (for example, ZOO Project has this kind of architecture), then following problems arise – only local WPS-services are available and scenarios and WPS-services are executed as processes of web-server (as web-server's CGI script), that serves WPS-application, so there are some performance limitations as well. Also such architecture results in significant timeouts and connection fault between web-server and client browser when long-lasting process is executed. The special component, written in C++ with integrated JavaScript interpreter Google V8, was developed for processing scenarios and executing WPS-services. Google V8 JavaScript interpreter allows distributed WPS-services to be integrated in the single JavaScript program. Google V8 compiles provided JavaScript down to the machine code and executes it, so it leads to significant performance. When the interpreter had being developed, programmers focused on stability and security. The programming interface of the JavaScript interpreter has only required for its purposes methods and is fully-controlled by Geoportal security policies. Basically the interpreter is separate virtual machine, whose access to remote resources is regulated. Control of WPS-services and intermediate data access was realized as well.

The programming interface of the interpreter is described below. Following methods are already realized:

1. CallWPS method – accepts parameters of WPS-application, description of WPS-service and list of parameters as an input in the form of JavaScript Objects. Then it forms a execution string and performs it, using CURL library for querying remote hosts. It supports long-lasting execution.

2. Matrix processing library – provided set of JavaScript functions, that allow create matrixes as a JavaScript Object. In fact these objects are created by special C++ library, but Google V8 allows to represent C++ structures as a JavaScript Objects, so simplicity of creating JavaScript code can be combined with C++ performance.

JavaScript wrapper is generated for every registered WPS-service and scenario, which accepts input parameters, processes them and launches the service. Wrappers are used for calling any WPS-service in the body of another service or scenario. Wrappers call described above method CallWPS at the service or scenario execution.

Separate server process for the interpreter is launched for every WPS scenario execution. Execution log, standard output streams are captured and stored in the database by the environment of distributed WPS-services for usability, debugging and quality assurance purposes.

**General environment architecture**

General structure of the environment of distributed WPS-services consists of the following subsystems:

– WPS-service catalog, intended to perform search, registration and launching of services and scenarios;

– WPS-service execution subsystem, that performs general control over the execution of service of scenario;

– WPS-scenario interpreter;

– WPS-service packages (local and remote).

## Testing

As an example, the road_pnt_pol service will be described. This scenario calculates pollution of point sources and roads in cells of the regular grid. The analyzed scenario uses several distributed WPS-services – vector2grid, road2grid, g_sum.

```javascript
function road_pnt_pol(housefile, roadsource, commonresult, sumpol){
    var houseresult='/tmp/hr.tif';
    var roadResult ='/tmp/hr.tif';
    vector2grid(housefile, houseresult);
    road2grid(roadsource, roadResult, sumpol);
    g_sum(roadResult, houseresult, commonresult);
    return true;
}
```

WPS-service vector2grid calculates pollution from point sources. Vector points file, that describes the location and amount of emitted pollution, serves as an input for this service. GeoTIFF file is created as a result of vector2grid execution, it displays summarized pollution in every cell of the grid. WPS-service road2grid calculates pollution from the roads. Vector file, that describes location of roads, theirs average load and summarized amount of pollution for the settlement. Result of road2grid is the GeoTIFF file, that describes summarized pollution in every cell. WPS-service g_sum takes two GeoTIFF's, that were generated as a result of execution above and combines them in the new file.

In this example, component of the environment of distributed WPS-services ensure following:

- scenario execution;
- passing remote WPS-services file data from the storage system, defined as input parameters;
- receive of resulting data and placing it in the storage system using predefined pathes;
- execution control, display of WPS-services messages;

Automation of WPS-service application using JavaScript shortened time of the calculation for the variety of air quality improvements, minimized number of user errors.

## Conclusions

Combination of WPS-services, data storage and PostgreSQL subsystems of Geoportal allows to simplify processing of geospatial data in the Internet. In order to pass data to WPS-services user only has to upload it to the Geoportal using input/output data entering subsystem, file manager and FTPS server. WPS-service results are collected and placed at the Geoportal as well and are available for download.

Variety of input widgets allows to define additional requirements for WPS-services input parameters, which are checked and filtered at the user-input stage. WPS-services can be provided with the set of widgets for its special parameters. Besides, convenient and simple interface is generated by widgets when service is executed.

The architecture of the environment of distributed WPS-services allows to extend its functionality with additional WPS-services. In the future the catalog of WPS-services will serve as an open bank of WPS-services, that solve problems from different fields. Ability to integrate methods using JavaScript allows to develop new methods of performing interdisciplinary researches.

## Acknowledgements

Development of the environment of distributed WPS-services was supported by RFBR 13-07-12080 (ofi_m), program of Presidium of RAS "Fundamental problems of mathematical modeling".